\documentclass[sigconf]{acmart} 

\usepackage{multirow}
\usepackage{xcolor, colortbl}
\usepackage{url}
\usepackage{hyperref}
\usepackage[many]{tcolorbox}    	
\usepackage{siunitx}

\newtcolorbox[auto counter]{somebox}[1][]{arc=5pt,auto outer arc,left=1pt,boxsep=0.5pt,boxrule=0.5pt,width=\columnwidth,right=1pt, #1}

\definecolor{main}{HTML}{5989cf}    
\definecolor{sub}{HTML}{D3D3D3}     

\newtcolorbox{boxB}{
    fontupper = \bf\color{main}, 
    boxrule = 1.5pt,
    colframe = main,
    rounded corners,
    arc = 5pt   
}

\newtcolorbox{boxC}{
    colback = sub, 
    boxrule = 0.5pt,  
    arc = 3pt   
}

\AtBeginDocument{%
  \providecommand\BibTeX{{%
    \normalfont B\kern-0.5em{\scshape i\kern-0.25em b}\kern-0.8em\TeX}}}

\setcopyright{rightsretained}
\copyrightyear{2024}
\acmYear{2024}

\acmConference[MSR '24]{21st International Conference on Mining Software Repositories}{April 15--16, 2024}{Lisbon, Portugal}
\acmBooktitle{21st International Conference on Mining Software Repositories (MSR '24), April 15--16, 2024, Lisbon, Portugal}\acmDOI{10.1145/3643991.3644880}
\acmISBN{979-8-4007-0587-8/24/04}

\begin{document}

\title{SATDAUG - A Balanced and Augmented Dataset for Detecting Self-Admitted Technical Debt}

\author{Edi Sutoyo}
\email{e.sutoyo@rug.nl}
\affiliation{%
  \institution{Bernoulli Institute, University of Groningen}
  \streetaddress{Nijenborgh 9}
  \city{Groningen}
  \country{Netherlands}
  \postcode{9747 AG}
}

\author{Andrea Capiluppi}
\email{a.capiluppi@rug.nl}
\affiliation{%
  \institution{Bernoulli Institute, University of Groningen}
  \streetaddress{Nijenborgh 9}
  \city{Groningen}
  \country{Netherlands}
  \postcode{9747 AG}
}

\renewcommand{\shortauthors}{Sutoyo and Capiluppi}

\begin{abstract}
Self-admitted technical debt (SATD) refers to a form of technical debt in which developers explicitly acknowledge and document the existence of technical shortcuts, workarounds, or temporary solutions within the codebase. Over recent years, researchers have manually labeled datasets derived from various software development artifacts: source code comments, messages from the issue tracker and pull request sections, and commit messages. These datasets are designed for training, evaluation, performance validation, and improvement of machine learning and deep learning models to accurately identify SATD instances. However, class imbalance poses a serious challenge across all the existing datasets, particularly when researchers are interested in categorizing the specific types of SATD. In order to address the scarcity of labeled data for SATD \textit{identification} (i.e., whether an instance is SATD or not) and \textit{categorization} (i.e., which type of SATD is being classified) in existing datasets, we share the \textit{SATDAUG} dataset, an augmented version of existing SATD datasets, including source code comments, issue tracker, pull requests, and commit messages. These augmented datasets have been balanced in relation to the available artifacts and provide a much richer source of labeled data for training machine learning or deep learning models.

  %

\end{abstract}

\maketitle

\section{Introduction}
Technical debt (TD), a metaphor coined by Cunningham \cite{cunningham1992wycash} in 1992, refers to shortcuts or workarounds taken during software development that result in future maintenance or modification efforts. Among the various forms of technical debt, self-admitted technical debt (SATD) stands as a distinctive category: SATD is a type of technical debt where developers explicitly acknowledge and document the existence of technical shortcuts, workarounds, or temporary solutions within the codebase \cite{potdar2014exploratory}. Successfully isolating SATD has demonstrated its significance, as it provides a complementary approach to static code analysis \cite{da2017using}. This is particularly important in the context of identifying specific SATD types or labels from other artifacts apart from the source code~\cite{sierra2019survey}.


In general, SATD takes the form of comments or annotations within the code, providing valuable insights into the developer's intentions, the potential risks associated with the technical shortcut, and the motivations behind the decision. Initially, Potdar and Shihab \cite{potdar2014exploratory} extracted the source code comments from four projects and analyzed 101,762 comments, identifying 62 patterns that are associated with SATD. Subsequently, Maldonado et al. \cite{da2017using} employed natural language processing (NLP) to identify SATD and made a breakthrough by providing a SATD dataset\footnote{https://github.com/maldonado/tse.satd.data} derived from the source code comments of 10 open-source projects. As a result, several researchers utilized this dataset extensively by proposing different approaches to identify SATD by employing contextualized patterns \cite{de2020identifying}, text mining \cite{huang2018identifying}, machine learning (ML) \cite{sala2021debthunter, chen2021multiclass}, and deep learning (DL) \cite{ren2019neural, santos2020self, zhu2023scgru}. 

In addition, researchers have been also keen on exploring additional artifacts, such as issue tracking systems \cite{dai2017detecting, xavier2020beyond, li2022identifying}, to identify SATD. In some cases, the authors made their datasets publicly available\footnote{https://zenodo.org/records/3701471}\textsuperscript{,}\footnote{https://github.com/yikun-li/satd-issue-tracker-data}. It is worth noting that all studies and datasets mentioned earlier have exclusively relied on a single artifact.

In the most recent attempts, authors have attempted to \textit{identify} SATD and \textit{categorize} it: rather than using only a single artifact, Li et al. \cite{li2023automatic} proposed an integrated approach for automatically identifying and categorizing SATD, and using four artifacts\footnote{https://github.com/yikun-li/satd-different-sources-data}: source code comments (CC), issue tracker sections (IS), pull request sections (PS), and commit messages (CM). 

Whether the SATD dataset is derived from a single or multiple artifacts, all the current datasets encounter a similar class imbalance issue. This is particularly evident when categorizing types of SATD (e.g., `design debt' or `test debt'): this could be due to a variety of factors, including the time and effort required to collect the essential data. The class imbalance problem can have a substantial impact on the identification and categorization of SATD items. Additionally, ML or DL models face difficulties in accurately categorizing SATD types, since they have to rely on limited data to learn semantic information. The issue of highly imbalanced data is a common obstacle that must be overcome in identifying SATD, particularly in categorizing types of SATD \cite{sridharan2021data}.


To address the current limitations of the existing dataset and research on SATD identification and categorization, we present \textit{SATDAUG}, an augmented and balanced dataset derived from several artifacts, i.e., source code comments (CC), issue tracker sections (IS), pull request sections (PS), and commit messages (CM). \textit{SATDAUG} originates directly from the dataset provided by Li et al. \cite{li2023automatic}. 

This paper is articulated as follows: Section 2 summarizes the motivation and the works related to this study. Section 3 presents the contributions of this paper, and Section 4 concludes the study.

\section{Related Work and Motivation}
To better understand and identify SATD automatically, Li et al. \cite{li2023automatic} developed a comprehensive dataset consisting of source code comments (CC), issue tracker sections (IS), pull request sections (PS), and commit messages (CM) from a variety of Apache open-source projects. This dataset includes 5,000 commit messages and 5,000 pull request sections from 103 Apache projects, each of which was manually labeled as either Not-SATD or SATD, including specific SATD types. Additionally, they created a larger dataset encompassing 23.7 million source code comments, 1.3 million commit messages, 0.3 million pull requests, and 0.6 million issues from the same 103 Apache projects. The Cohen's kappa coefficient \cite{landis1977measurement} was employed to assess the bias risk and reliability among authors during the manual labeling of the datasets. The outcome revealed `substantial agreement' among all authors, with a Cohen's kappa coefficient of +0.74. This means that all three authors were in high agreement with each other, and according to Cohen's kappa coefficient \cite{landis1977measurement}, their ratings can be considered to be reliable.

In their seminal work, Maldonado et al. \cite{da2017using} noted that their SATD dataset, specifically the one relative to the source code comments, does not differentiate between `code debt' and `design debt', as they are highly similar. Being a superset of Maldonado et al. \cite{da2017using} dataset, also the dataset provided by \cite{li2023automatic} combines the two types of SATD, and names it `code/design debt'. The resulting four datasets consist of four distinct types of SATD: i) code/design debt (C/D), ii) documentation debt (DOC), iii) test debt (TES), and iv) requirement debt (REQ). These four types of SATD have been frequently used to categorize SATD in the various software development artifacts \cite{da2017using, li2022identifying}. Table~\ref{tab:tb_dataset} shows the number of specific types of SATD that have been shared in this dataset.

\begin{table}[htpb]
    \small
    \caption{Dataset from Li et al. \cite{li2023automatic}} 
    \begin{tabular}{p{1.5cm} p{1.2cm} p{1.4cm} p{1.3cm} p{1.3cm}}
        \toprule
        Labels & CC & IS & PS & CM\\
        \midrule
        
        C/D & 2,703 & 2,169 & 510 & 522\\
        DOC & 54 & 487 & 101 & 98\\
        TES & 85 & 338 & 68 & 58\\
        REQ & 757 & 97 & 20 & 27\\\hline
        SATD & 3,599 & 3,091 & 699 & 705\\
        Not-SATD & \textbf{58,676} & \textbf{20,089 }& \textbf{4,301} & \textbf{4,295}\\
        \bottomrule
    \end{tabular}
    \label{tab:tb_dataset}
\end{table}

The analysis of the dataset from Table~\ref{tab:tb_dataset} shows that the distribution of specific types of SATD is highly imbalanced. As an example, the number of "DOC debt" instances extracted from the CC artifact has only 54 rows of usable data: this compares to only 2\% of the data in the "C/D debt" and a mere 1\% compared to "Not-SATD". The other three datasets, derived from IS, PS, and CM artifacts, also exhibit a significant class imbalance.

To solve the problem of the imbalanced class, we utilized AugGPT \cite{dai2023chataug}, a text augmentation technique based on the ChatGPT language model. The primary objective was to generate \textit{paraphrased}, additional versions of existing texts, while maintaining their original meaning. It is important to preserve the key semantics of the text while generating paraphrased versions \cite{xie2022multi}: any unintended alterations to the data can potentially lead to mislabeling of the texts. 



\section{Contributions of the paper} 
In this section, we cover the following aspects: we detail the \textit{methodology} employed (Section~\ref{lbl:_methodology}), we give a \textit{description of the dataset and file formats} (Section~\ref{lbl:_description}); we highlight our \textit{previously study} that utilized such approach (Section~\ref{lbl:_dataset_used}); we discuss the \textit{originality of the dataset} (Section~\ref{lbl:_originality}). Additionally, ideas for \textit{further improvements} that could be made to the dataset are explained in Section~\ref{lbl:_improvement}, while insights into \textit{prospective research} that the dataset could address are discussed in Section~\ref{lbl:_potential_research}. Finally, an exploration of the \textit{limitations and challenges} encountered during the utilization of the dataset is presented in Section~\ref{lbl:_limitation}.

\subsection{Methodology: Data Augmentation}
\label{lbl:_methodology}
\begin{figure}[htbp!]
\centerline{\includegraphics[width=240px]{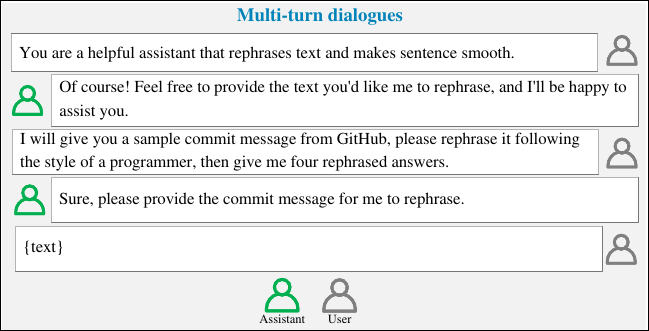}}
\caption{Multi-turn dialogues prompt for CM artifact}
\label{fig_dialogue}
\end{figure}

Dai et al. \cite{dai2023chataug} identified two design prompt alternatives for data augmentation: single-turn and multi-turn dialogues. A `multi-turn dialogue prompt' refers to a series of conversational exchanges (or turns) provided as input, and it involves multiple sequential messages that simulate a dialogue. In this study, we adopted the multi-turn prompt dialogue approach, guided by the principles outlined in their work. Additionally, we incorporated context and persona to ensure the generated augmentation data closely resembled the original text \cite{white2023prompt}. An example of a multi-turn prompt dialogue for augmenting the CM instances used in this study is shown in Figure~\ref{fig_dialogue}.

Table~\ref{tab:tb_sample} presents a sample message generated from one of the datasets employed in this research, namely the commit message (CM) artifact. The left column shows the original message and the right column displays the same message after undergoing four rounds of paraphrasing.

\begin{table}[htpb]
    \centering
    \small
    \caption{Generated samples on CM artifact} 
    \begin{tabular}{p{1.5cm} p{6.2cm}}
        \toprule
        Original & Generated\\
        \midrule
        \multirow{4}{6em}{CAMEL-8023: Polished the java doc} & 1. CAMEL-8023: Enhanced the JavaDoc\\
        & 2. CAMEL-8023: Improved the JavaDoc documentation\\
        & 3. CAMEL-8023: Fine-tuned the JavaDoc\\
        & 4. CAMEL-8023: Revised and improved the JavaDoc\\
        \bottomrule
    \end{tabular}
    \label{tab:tb_sample}
\end{table}



For a more detailed explanation, let's consider an example from the original, pre-augmented dataset and focus on the commit message (CM) artifact (last column of Table~\ref{tab:tb_dataset}). As shown in Table~\ref{tab:tb_augmentation_process}, each specific type of SATD (i.e., C/D, DOC, TES, and REQ debt) underwent varying degrees of data augmentation. The goal was to achieve a balanced or approximately equivalent number of instances for each SATD type, with reference to the highest number of instances found in the C/D debt label (which totaled 522 instances).

\begin{table}[htpb]
    \small
    \caption{Dataset from CM artifact} 
    \begin{tabular}{p{1.5cm} p{1.5cm} p{2.5cm} p{1.3cm}}
        \toprule
        SATD Type & Original amount & \# of times each instance is augmented & Final amount\\
        \midrule
        C/D & 522 & - & 522\\
        DOC & 98 & 4 & 490\\
        TES & 58 & 8 & 522\\
        REQ & 27 & 18 & 513\\

        \bottomrule
    \end{tabular}
    \label{tab:tb_augmentation_process}
\end{table}

In the case of DOC debt, there were initially only 98 data points: in order to increase this amount to approximate the quantity of C/D debt data (i.e., 522 instances), \textit{each} of these instances was augmented four times, thus yielding a total of 490 instances. This was facilitated through a multi-turn prompt dialogue (as shown in Figure~\ref{fig_dialogue}). 

The same augmentation process was employed for augmenting other SATD types, namely TES and REQ debt, but with different multipliers. Each TES debt instance underwent data augmentation 8 times, while REQ debt underwent it 18 times. As a result, the TES debt now comprises 522 instances, while the REQ debt contains 513 instances. The resulting augmented dataset is demonstrated in Table~\ref{tab:tb_augmented_dataset}.  

In order to measure the similarity between the original and augmented texts, cosine similarity \cite{wang2020measurement} is employed. By using BERT Embeddings, the cosine similarities for CC, IS, PL, and CM are 0.864, 0.878, 0.862, and 0.748, respectively. A high similarity value indicates that the generated texts have high faithfulness and compactness, and are close to the original text \cite{dai2023chataug}.

\begin{table}[htpb]
    \small
    \caption{Augmented dataset for each label} 
    \begin{tabular}{p{1.5cm} p{1.2cm} p{1cm} p{1.3cm} p{1cm}}
        \toprule

        SATD Type & CC & IS & PS & CM\\
        \midrule
        C/D & 2,703 & 2,169 & 510 & 522\\
        DOC & 2,700 & 1,948 & 505 & 490\\
        TES & 2,635 & 2,028 & 476 & 522\\
        REQ & 2,271 & 2,134 & 500 & 513\\

        \bottomrule
    \end{tabular}
    \label{tab:tb_augmented_dataset}
\end{table}

Table~\ref{tab:tb_shannon} shows the results of the Shannon entropy \cite{shannon2001mathematical} as the measurement of the class distribution in the datasets. A higher entropy value indicates a more balanced distribution, while a lower value denotes a more imbalanced distribution. This table presents the entropy scores for the original and augmented datasets, grouped based on the identification and categorization of SATD tasks. The table clearly shows that the balance of the datasets for both SATD Identification and Categorization has improved.

\begin{table}[htpb]
    \small
    \caption{Imbalance of classes using Shannon Entropy \cite{shannon2001mathematical}}  
    \begin{tabular}{p{1.1cm} p{1.1cm}|p{1.6cm} p{1.1cm}|p{1.2cm}}
        \toprule
        \multirow{2}{5em}{Artifacts} & \multicolumn{2}{c}{\textbf{Identification}} & \multicolumn{2}{c}{\textbf{Categorization}}\\
        \cline{2-5}
        & Original & Augmented & Original & Augmented\\
        \midrule
        CC & 0.231 & 0.611 & 0.500 & 0.998\\
        IS & 0.569 & 0.873 & 0.642 & 0.999\\
        PS & 0.585 & 0.901 & 0.604 & 0.999\\
        CM & 0.589 & 0.909 & 0.596 & 0.999\\
        \bottomrule
    \end{tabular}
    \label{tab:tb_shannon}
\end{table}

\subsection{Dataset Description}
\label{lbl:_description}


In accordance with the original dataset, the \textit{SATDAUG} dataset comprises four distinct CSV files delineated by the artifacts under consideration in this study. Each CSV file encompasses a text column and a class, which indicate classifications denoting specific types of SATD or Not-SATD, and also a status indicating whether the text is original or augmented. This \textit{SATDAUG} dataset generated using AugGPT along with the replication package is publicly accessible\footnote{https://zenodo.org/records/10521909}. 











\subsection{Previous Uses of the SATDAUG Dataset}
\label{lbl:_dataset_used}
In a previous study~\cite{sutoyo2023twostep}, we utilized the augmented dataset to train two deep learning models (BiLSTM and BERT) for SATD identification and categorization, respectively. Our initial findings indicate that the augmented datasets lead to greatly enhanced generalization and improved model performance compared to the original dataset. To make this paper self-contained, the performance results in terms of the F1-score of our approach are presented in Tables ~\ref{tab:tb_bilstm_result} and ~\ref{tab:tb_bert_result}. In brackets, we also summarize the F1-scores we obtained in the identification and categorization phases, but when no augmentation was performed on the datasets.

\begin{table}[htpb]
    \small
    \caption{SATD identification using BiLSTM+AugGPT} 
    \begin{tabular}{p{1.2cm} p{2.0cm} p{2.0cm}}
        \toprule
        \multirow{2}{6em}{Artifacts} & \multicolumn{2}{c}{F1-score}\\
        \cline{2-3}
        &Not-SATD&SATD\\
        \midrule
        CC
        & 0.952 (\textit{0.952}) & 0.947 (\textit{0.799})\\
        
        IS
        & 0.937 (\textit{0.937}) & 0.820 (\textit{0.559})\\
    
        PS
        & 0.917 (\textit{0.915}) & 0.806 (\textit{0.422}) \\
        
        CM
        & 0.940 (\textit{0.926}) & 0.861 (\textit{0.668}) \\

        \bottomrule
    \end{tabular}
    \label{tab:tb_bilstm_result}
\end{table}

\begin{table}[htpb]
    \small
    \caption{SATD categorization using BERT+AugGPT} 
    \begin{tabular}{p{0.9cm} p{1.45cm} p{1.45cm} p{1.45cm} p{1.45cm}}
        \toprule
        \multirow{2}{7em}{Artifacts} & \multicolumn{4}{c}{F1-score}\\
        \cline{2-5}
         & C/D & DOC & TES & REQ \\
        

        

        
        \midrule
        CC & 0.857 (\textit{0.725}) & 0.965 (\textit{0.626}) & 0.965 (\textit{0.540}) & 0.796 (\textit{0.585}) \\
        IS & 0.899 (\textit{0.486}) & 0.922 (\textit{0.457}) & 0.922 (\textit{0.432}) & 0.951 (\textit{0.437}) \\
        PS & 0.865 (\textit{0.539}) & 0.895 (\textit{0.441}) & 0.951 (\textit{0.461}) & 0.942 (\textit{0.325}) \\
        CM & 0.882 (\textit{0.536}) & 0.926 (\textit{0.659}) & 0.941 (\textit{0.449}) & 0.940 (\textit{0.255}) \\
        \bottomrule
    \end{tabular}
    \label{tab:tb_bert_result}
\end{table}


\subsection{Originality of the Dataset}
\label{lbl:_originality}

Based on our extensive review of the literature \cite{sutoyo2023detecting}, spanning the last decade, at least 34 studies have been conducted with a focus on SATD \textit{detection} within source code comments and directly making use of the dataset provided by Maldonado et al. \cite{da2017using}. While some studies concentrate solely on identifying SATD and Not-SATD instances \cite{huang2018identifying, flisar2018enhanced, ren2019neural}, more sophisticated investigations have attempted to categorize diverse types of technical debt, such as `design' and `requirement' debt \cite{wattanakriengkrai2018identifying, santos2020self}. Further advanced studies endeavor to specifically \textit{categorize} specific types of SATD, including design, requirement, and defect debt \cite{zhu2021detecting}. Additionally, certain inquiries extend their scope to encompass several types of SATD, including defect, test, documentation, design, and requirement debt \cite{di2022pilot, zhu2023scgru}. 


The originality of \textit{SATDAUG} is rooted in the augmented number of examples in the classes that are seriously imbalanced. This augmented dataset is expected to contribute to research on SATD identification and categorization: we believe that \textit{SATDAUG} has the potential to further enhance (or even revise) the findings, including the precision, recall, and F1-score, of all of the previous studies that focused on the categorization of SATD types \cite{wattanakriengkrai2018identifying, santos2020self, sala2021debthunter, zhu2021detecting, chen2021multiclass, zhu2023scgru}.

\subsection{Further Improvements}
\label{lbl:_improvement}
There are two improvements that we are considering for this dataset: first, we will augment the PS and CM artifacts (last two columns of Table~\ref{tab:tb_augmented_dataset}) by aligning the quantity of data more closely with that of the CC and IS artifacts (first two columns of Table~\ref{tab:tb_augmented_dataset}) in the future. However, a challenge arises with repeated augmentation, as it could lead to redundancy and noise due to the high similarity in texts. Therefore, our approach must be accompanied by exploring the impact of re-augmentation on the robustness and generalization capabilities of the models. A comprehensive investigation could involve varying degrees of re-augmentation and assessing how it influences model performance, including metric evaluation results.

The second area for improvement concerns the quality of the augmentation. It is crucial to not only consider paraphrases that are semantically similar but also to ensure the textual diversity of the paraphrasing generated. Moreover, more diverse generated results are expected for the models to achieve stronger generalization and prevent overfitting, even with less semantic similarity \cite{yang2020generative}.




\subsection{Potential Research Applications} 
\label{lbl:_potential_research}
Below are several research opportunities that can be examined using the \textit{SATDAUG} dataset.

\subsubsection{Improvement of performance in SATD identification and categorization:} Imbalanced data poses a common challenge in classification tasks, which also occurs in tasks involving SATD identification. 
Models trained on imbalanced datasets often encounter difficulties in generalizing patterns from minority classes to new data. This has been observed in numerous prior studies, where ML/DL models face challenges in accurately categorizing specific types of SATD \cite{da2017using, chen2021multiclass, li2022identifying, li2023automatic, sala2021debthunter, santos2020self, zhu2023scgru}. We aim to motivate both practitioners and researchers to adopt our dataset as a benchmark for SATD identification. As the training data for identifying and categorizing SATD increases, it is expected that using this larger dataset will improve the performance of identifying SATD, and also when categorizing specific types of SATD. This is because the effectiveness of ML/DL models often sees improvement in conjunction with the size of the dataset employed \cite{zhu2016we}. 

\subsubsection{Rerun previous studies on SATD identification and categorization:}
For nearly a decade, SATD research has significantly depended on the dataset provided by Maldonado et al.\cite{da2017using}, even though the dataset is highly imbalanced. Imbalanced data consistently hinders the effectiveness of models, as it often results in the minority class being missed \cite{fernandez2018learning}. 
We see the need for replicating and rerunning past studies on identifying and categorizing SATD using this augmented and balanced dataset. Comparing the results with the original can provide invaluable insights and perhaps revise the previous results.


\subsubsection{Improved identification approaches for better SATD management}
Managing technical debt begins with the explicit identification and categorization of such debt \cite{li2015systematic}, so it is critical that developers are aware of the presence of technical debt items. By improving the identification and categorization performance, we can reduce the likelihood of overlooking certain SATD items; help developers reduce software maintenance costs; and enhance the quality of TD repayment and corrective actions (e.g., TD removal).

\subsection{Limitations}
\label{lbl:_limitation}
We are aware of one limitation of our dataset: the augmentation process relies heavily on the labels assigned from the manual labeling of each data point in the original dataset provided by Li et al. \cite{li2023automatic}. Since the number of times each instance is augmented, for each specific type of SATD, depends on those original datasets, any wrong labels from the original labelling process will consequently be expanded into the augmentation dataset.

Take for instance, a single REQ debt instance from the CM artifact being augmented 18 times: if one data point was mislabeled, that would result in an additional 18 instances that are also mislabeled in the augmented dataset. This will inevitably impact the model's performance as it incorporates inaccurately labeled data during the training process. We can only rely on the accuracy of the labeling from the previous researchers, but a more in-depth analysis (e.g., random sampling and manual annotation of the augmented dataset) could be run to mitigate this limitation.

\section{Conclusion}
In this study, we presented \textit{SATDAUG}, an augmented dataset created by balancing the existing SATD dataset, and using an automated augmentation approach. Developing this SATD augmented dataset from several artifacts has the potential to enhance the training datasets for SATD identification, capturing a broader range of variations that might be encountered in real-world scenarios. 

This augmented dataset not only enhances the performance of the model but also helps mitigate bias, especially in cases where ML or DL models are trained on datasets that do not accurately represent the real world. By augmenting training data with real-world variations, models are expected to learn to generalize better to new situations and reduce bias, leading to more accurate and unbiased SATD identification and categorization.

\balance

\bibliographystyle{ACM-Reference-Format}
\bibliography{Main}











\end{document}